\documentclass{article}

\usepackage{arxiv}

\usepackage[utf8]{inputenc} 
\usepackage[T1]{fontenc}    
\usepackage{hyperref}       
\usepackage{url}            
\usepackage{booktabs}       
\usepackage{amsfonts}       
\usepackage{amsmath}        
\usepackage{amssymb}        
\usepackage{nicefrac}       
\usepackage{microtype}      
\usepackage{graphicx}
\usepackage{natbib}
\usepackage{doi}
\usepackage{xspace}         
\usepackage{paralist}       
\usepackage[caption=false]{subfig}  
\usepackage{xcolor}         

\graphicspath{{figures/}}

\newcommand{\cGAW}{cGAW\xspace}
\newcommand{\GAW}{GAW\xspace}

\title{Estimating Absolute Web Crawl Coverage from Longitudinal Set Intersections}

\author{
	Michael Paris\thanks{Corresponding author} \\
	Common Crawl Foundation\\
	\texttt{micha@commoncrawl.org}
	\and
	Grigori Paris\\
	Independent Researcher\\
	\texttt{grigori.paris.berlin@gmail.com}
	\and
	Fabian Baumann\\
	University of Pennsylvania\\
	\texttt{baumannf@sas.upenn.edu}
	\\
}

\renewcommand{\shorttitle}{Estimating Absolute Web Crawl Coverage}

\hypersetup{
pdftitle={Estimating Absolute Web Crawl Coverage from Longitudinal Set Intersections},
pdfsubject={cs.IR, cs.DL},
pdfauthor={Michael Paris},
pdfkeywords={web archive, crawl coverage, set intersection, longitudinal analysis, sampling fraction},
}

\begin{document}
\maketitle

\begin{abstract}
Web archives preserve portions of the web, but quantifying their completeness remains challenging.
Prior approaches have estimated the coverage of a crawl by either comparing the outcomes of multiple crawlers, or by comparing the results of a single crawl to external ground truth datasets.
We propose a method to estimate the absolute coverage of a crawl using only the archive's own longitudinal data, i.e., the data collected by multiple subsequent crawls.
{Our key insight is that coverage can be estimated from the {empirical} URL overlaps between subsequent crawls, which are in turn well described by a simple urn process. The parameters of the urn model can then be inferred from longitudinal crawl data using linear regression.}
{Applied to our focused crawl configuration of the German Academic Web (15 semi-annual crawls between 2013--2021), we find a coverage of approximately 46\% of the crawlable URL space for the stable crawl configuration regime.}
Our method is extremely simple, requires no external ground truth, and generalizes to any longitudinal focused crawl. 
\end{abstract}

\section{Introduction}
\label{sec:intro}

Focused web crawls---archives targeting specific domains like academic institutions, government websites, or cultural heritage---must terminate at some point.
The German Academic Web (\GAW), for instance, crawls approximately 150 German university seeds semi-annually, stopping after collecting roughly 100 million records.
This raises the question: how much of the \emph{crawlable} German Academic Web (\cGAW) is archived by the \GAW?

A web crawl is defined by its crawl configuration, which specifies the heuristics and rules to collect websites. For instance, it defines the termination threshold $M$, i.e., the total number of URLs to be collected in the crawl. What remains unknown, however, is the set of URLs ($N$) that could, in principle, be collected by a given crawler and its crawl configuration. As a result we are faced with the fundamental limitation of a single, {isolated} crawl: it does not allow us to determine its coverage $c=M/N$ of the underlying web, i.e. it remains unknown how much of the crawlable web space does it actually capture.


Several strategies have addressed related problems. \citet{bharat1998technique} estimated \emph{relative} search engine sizes by comparing URL sets across multiple engines.
Their method yields ratios like ``Engine A is 2.3$\times$ larger than Engine B'' but cannot determine what fraction of the total web either captures. Entity-based approaches use external ground truth (e.g., named entities from reference databases) to measure recall of known items.
This requires domain-specific datasets and measures coverage only for that entity class, not the full URL space. Quality frameworks like SHARC \citep{denev2011sharc} assess temporal coherence and resource completeness of archived pages, but not overall domain coverage. 

Here we provide an interpretable method to estimate absolute coverage from longitudinal crawl data alone.
The contribution is twofold. First, we gain theoretical insight by modeling the crawling process using a simple urn model. And second, we estimate coverage by fitting the model the longitudinal crawl data.

The remainder of the paper is structured as follows. Section~\ref{sec:related} reviews related work on set similarity measures and coverage estimation.
Section~\ref{sec:method} formalizes the random sampling model and derives the self-intersection interpretation.
Section~\ref{sec:results} applies our method to the German Academic Web, showing the heatmap, observed decay patterns, and coverage estimates.
Section~\ref{sec:discussion} discusses implications and limitations.

\begin{table}[t]
\centering
\caption{Positioning of our approach relative to prior work.}
\label{tab:positioning}
\begin{tabular}{lll}
\toprule
\textbf{Method} & \textbf{Output} & \textbf{Requirements} \\
\midrule
Bharat \& Broder (1998) & Relative: $\frac{\text{Size}(A)}{\text{Size}(B)}$ & Multiple crawlers \\
Paris \& J\"aschke (2020) & Recall of entity class & External ground truth \\
\textbf{This paper} & \textbf{Absolute: $c = M/N$} & \textbf{Only longitudinal data} \\
\bottomrule
\end{tabular}
\end{table}

\section{Related Work}
\label{sec:related}

Estimating an unknown population size from overlapping samples dates to \citet{Petersen1896} and \citet{lincoln1930calculating}, who formalized the estimator $\hat{N} = n_1 n_2 / m$ for wildlife abundance, later bias-corrected by \citet{Chapman1951}.
\citet{Seber1982} provides the standard reference covering closed- and open-population models, including the Jolly--Seber model for populations with births and deaths.
\citet{Fienberg1972} reconceptualized multiple-recapture as incomplete contingency tables, enabling heterogeneous-catchability extensions.

These ecological methods were first applied to the web by \citet{Lawrence1998}, who used pairwise URL overlap across six search engines to estimate the indexable web size via Lincoln--Petersen.
They noted that positive dependence between engines inflates overlap, making their estimate a lower bound.
\citet{Lawrence1999} complemented this with random IP sampling, and \citet{bharat1998technique} derived relative size ratios from random-query sampling, yielding only relative sizes and requiring multiple engines.
\citet{Dobra2004} reanalyzed the data with Bayesian log-linear models incorporating Rasch-type heterogeneity, estimating roughly 2.5$\times$ the original figure and demonstrating that ignoring capture heterogeneity causes severe underestimation.

While these studies compared multiple engines at a single point in time, \citet{BarIlan1998,BarIlan1999} conducted the first longitudinal study, tracking a query across engines over five months and categorizing URLs as lost, dropped, forgotten, or recovered.
\citet{BarIlan2002methods} formalized measures for temporal self-overlap and relative coverage.

Central to all overlap-based approaches is the choice of similarity measure.
The symmetric Jaccard index \citep{jaccard1901etude} and Dice--S{\o}rensen coefficient \citep{dice1945measures,sorensen1948method} contrast with the asymmetric containment $C(A,B) = |A \cap B|/|A|$ introduced by \citet{broder1997resemblance}, who also proposed MinHash sketches for its efficient estimation \citep{Broder2000minwise}.
\citet{garg2015asymmetric} showed that asymmetric measures often outperform symmetric ones in retrieval tasks.

The probabilistic foundation comes from urn model theory: \citet{kalinka2013probability} showed that expected intersection sizes under independent sampling follow the hypergeometric distribution, with \citet{mahmoud2008polya} providing comprehensive treatment of P\'{o}lya urn models.
This connection to capture-recapture is established in ecology \citep{Seber1982} but has not previously been applied to web crawl coverage.

A key assumption of urn-based sampling is a stable population, yet the web undergoes continuous turnover.
\citet{koehler2002web,koehler2004longitudinal} found a web page half-life of approximately two years; \citet{gomes2006modeling} modeled URL persistence; \citet{fetterly2003evolution} and \citet{wren2008url} quantified change and decay rates at scale.

Complementary work on web archive quality---capture ``blur'' and coherence \citep{denev2011sharc}, missing embedded resources \citep{brunelle2015not}, and temporal inconsistency \citep{ainsworth2015temporal}---focuses on individual page fidelity rather than domain coverage.
As \citet{baack2024critical} found, even the Common Crawl team expressed fundamental uncertainty, with their director noting the web is ``practically infinite.''
Overall, existing work either (1)~uses entity-based proxies requiring external datasets \citep{paris2020assess}; (2)~compares multiple engines for only relative sizes \citep{bharat1998technique,Lawrence1998}; or (3)~studies temporal decay without connecting it to coverage \citep{koehler2002web,BarIlan1999}.
Our contribution is to extract absolute sampling fractions from longitudinal self-intersections of a single archive.
%

\section{Theoretical intuition and urn model}
\label{sec:method}

Let $u_t$ denote the set of URLs captured in a given crawl at time $t$, with $|u_t| = M$ being the number of URLs captured in the crawl. Furthermore, let $N$ denote the size of the crawlable web, which is represented by the size of the set of all URLs within the archive's scope that could potentially be harvested. Note that while we assume to know the value of $M$, we do not have access to the value of $N$, as a crawl will not capture all crawlable URLs. The ratio $c=M/N$ is defined as the coverage of the crawl.


The urn represents a population of $N$ unique URLs. Time advances in discrete steps $t=1,\dots,T$, which corresponds to successive crawls.
At each time step, the urn process consists of two steps: First, turnover occurs in the underlying population. This means that a fraction of $(1-\alpha)$ of the $N$ URLs is removed and replaced by the same number of newly introduced URLs, such that $N$ remains constant. In other words, the parameter $\alpha\in[0,1]$ captures the temporal persistence of URLs across two subsequent crawls. Second, conditional on the resulting population, a crawl sample $u_t$ is generated  by drawing $M$ URLs uniformly at random without replacement from the urn.

Crucially, this simple model allows us to derive an analytical expression for the overlap between two crawl at different times, $u_i$ and $u_j$. First, let us consider an arbitrary URL that appears in the crawl at time $t=1$. The probability that this URL survives the population turnover for $T-1$ steps is $\alpha^{T-1}$. Second, conditional on surviving until time $T$, the probability that it is included in the crawl at time $T$ corresponds to $M/N$, i.e., the uniform sampling probability from the urn. Since the initial crawl contains $M$ URLs, the expected size of the overlap between the first and the $T$-th crawl is therefore given by
\begin{equation}
    \mathbb{E}[|u_1 \cap u_T|] = M \cdot \alpha^{T-1} \cdot \frac{M}{N}
= \frac{M^2}{N}\,\alpha^{T-1}.
\end{equation}
Dividing by $M$, and substituting $c=M/N$, yields the expected fraction of URLs in the initial crawl that reappear at time $T$,
\begin{equation}\label{eq:model1}
f(T)
= c\,\alpha^{T-1}.
\end{equation} 
The important feature of Eq.~\ref{eq:model1} is that it can readily be used to fit the empirical observations of temporal crawl overlaps to extract the value of $c$. Interestingly, Eq.~\ref{eq:model1} is closely linked to previous work, specifically, Broder's containment measure, defined as \citep{broder1997resemblance}
\begin{equation}\label{eq:broder1}
	g(u_i, u_j) = \frac{|u_i \cap u_j|}{|u_i|}\,.
\end{equation}
As defined in Eq.~\ref{eq:broder1}, $g(u_i, u_j)$ denotes the fraction of crawl $u_i$ that also appears in crawl $u_j$, and has been used for cross-sectional multi-engine comparisons \cite{bharat1998technique}. In contrast to its original conception of $g$, our model allows to interpret Eq.~\ref{eq:broder1} as the definition of URL overlap between two crawls separated by $t=i-j$, hence 


\begin{equation}
	g^* = g(u_i, u_i),
\end{equation}

is interpreted as ``self-intersection'', which we recover from Eq.~\ref{eq:model1} by setting $T=1$ and thus get $g^*=c$. Equivalently, this result can be obtained directly from Eq.~\ref{eq:broder1} by assuming two 
independent draws ($i=j$) of size $M$ from a population of $N$ URLs, where the expected intersection size follows from the hypergeometric distribution \citep{kalinka2013probability}, i.e., we get\footnote{Note that the hypergeometric (sampling without replacement) and binomial (independent Bernoulli trials) models give identical expected intersections.
They differ only in variance---hypergeometric variance is smaller by the finite population correction $(N-M)/(N-1)$. For point estimation, the models are equivalent.}
\begin{equation}
	g^*=\mathbb{E}\!\left[\frac{|u_1 \cap u_T|}{M}\right]= \frac{1}{M} \frac{M^2}{N}=c\,.
\end{equation}

Importantly, this interpretation holds even if the crawl itself is non-uniform.
Consider a fixed crawl $u$ with $|u| = M$ obtained by any sampling method (uniform, breadth-first, preferential, etc.) from population $U$ with $|U| = N$.
If we compare against a uniform sample $\hat{u}$ from $U$, the expected intersection is:
\begin{equation}
    \mathbb{E}\!\left[\frac{|u \cap \hat{u}|}{M}\right] = \frac{M}{N} = c.
\end{equation}
This follows because each element of $u$ has probability $M/N$ of being sampled into $\hat{u}$, regardless of how $u$ was constructed.
Equivalently, a uniform sample of size $M$ would contain $c \cdot M$ elements in common with the crawl.

Thus the coverage estimate $c$ has three equivalent interpretations:
\begin{inparaenum}[(i)]
\item the fraction of the cGAW captured by the crawl,
\item the expected overlap fraction with a uniform sample, and
\item the normalized self-intersection under independent resampling.
\end{inparaenum}

Extending this framework to pairs of crawls, we compare the observed relative overlap $g(u_i, u_j) = |u_i \cap u_j|/M$ against the theoretical baseline of uniformly sampled counterparts, denoted $\hat{u}_i$ and $\hat{u}_j$.
Under the urn model with independent uniform sampling, the expected relative overlap is determined solely by the coverage $c$ and the temporal decay $\alpha$:
\begin{equation}
    \mathbb{E}[g(u_i, \hat{u}_j)] = \mathbb{E}[g(\hat{u}_i, \hat{u}_j)] = c \cdot \alpha^{|i-j|}
\end{equation}
However, real-world crawling processes often exhibit temporal correlations - for instance, through persistent seeds that force the crawler into the same local subgraph across repetitions. 
We model this deviation by introducing a \emph{crawler bias coefficient} $R_{ij}$, defined as the ratio between the actual and theoretical expectations:
\begin{equation}
    \label{eq:overlap_ratio}
    R_{ij} = \frac{\mathbb{E}[g(u_i, u_j)]}{\mathbb{E}[g(u_i, \hat{u}_j)]} = \frac{\mathbb{E}[g(u_i, u_j)]}{c \cdot \alpha^{|i-j|}}
\end{equation}

The overlap ratio $R_{ij}$ serves as a diagnostic metric for crawler behavior: $R_{ij} \approx 1$ implies a uniform process, while $R_{ij} > 1$ reveals positive correlation where the crawler systematically revisits specific subgraphs. 
This bias is empirically isolated as the residual term in the log-linear regression:
\begin{equation}
    \underbrace{\log R_{ij}}_{\text{Residual}} = \log g(u_i, u_j) - (\log c + |i-j| \log \alpha)
\end{equation}
\section{Results}
\label{sec:results}
    
We apply our methodology to the German Academic Web (\GAW), a longitudinal focused crawl of German university websites.
The part of the \GAW considered here consists of 15 semi-annual focused crawls conducted between December 2013 and December 2021.
Each crawl started from approximately 150 seed domains corresponding to German universities with doctorate-granting rights.
The Heritrix crawler~\cite{heritrix} was used with a breadth-first traversal policy.
Crawls terminate after collecting approximately 100 million records~\cite{paris2020assess}.

\begin{figure}[t]
	\centering
	\subfloat[Heatmap displays the containment between pairs of crawls.\label{fig:heatmap}]{%
		\includegraphics[width=0.48\columnwidth]{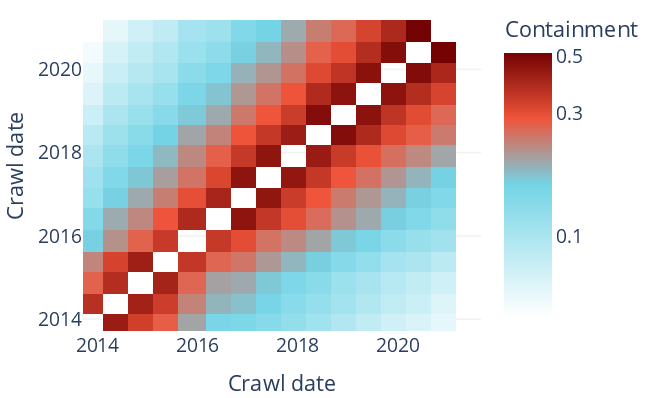}}
	\hfill
	\subfloat[Compressed representation of the left panel onto $\Delta t$.\label{fig:scatter}]{%
		\includegraphics[width=0.48\columnwidth]{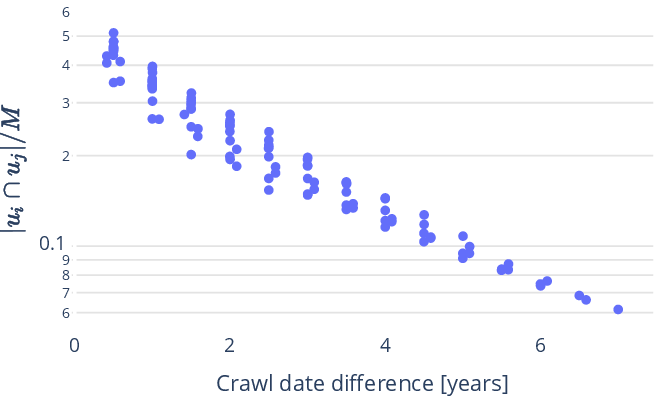}}
	\caption{Pairwise URL set intersections for the 15 \GAW crawls. (a) The heatmap reveals temporal decay structure; the diagonal shows estimated self-intersection values. (b) Log-transformed containment vs.\ time difference exhibits a linear relationship.}
	\label{fig:heatmap-scatter}
\end{figure}

The data of our longitudinal crawls of the \cGAW allows us to empirically observe the containment between two crawls $g(u_i, u_j)$ for $i>j$. The results are depicted in Fig.~\ref{fig:heatmap-scatter}, which shows two views of the pairwise containment measures.
Figure~\ref{fig:heatmap-scatter}(a) displays the $15 \times 15$ matrix of $g(u_i, u_j)$ values as a heatmap.
The off-diagonal entries show clear temporal structure, and values decrease with distance from the diagonal, i.e. for increasing time difference between two crawls, reflecting URL turnover.
This visualization makes explicit our estimation target, namely the hypothetical intersection of two independent samples drawn at the same moment, i.e., $\Delta t=0$. 
Figure~\ref{fig:heatmap-scatter}(b), depicts a scatter plot showing the log-transformed containment measure $\log g(u_i, u_j)$ as a function of the time difference between two crawls for every crawl pair.

\begin{figure}[t]
	\centering
	\subfloat[Empirical \GAW data.\label{fig:sim-empirical}]{%
		\includegraphics[width=0.48\columnwidth]{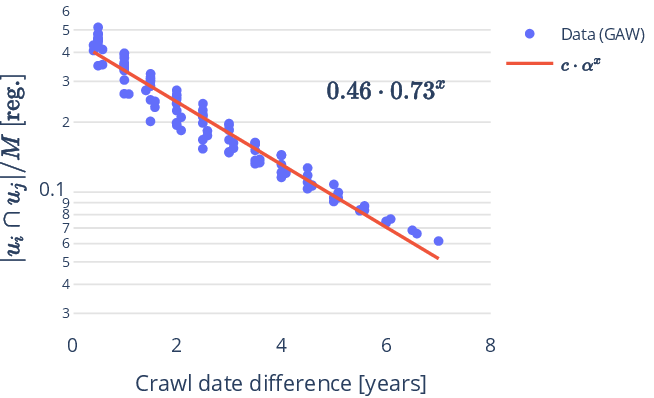}}
	\hfill
	\subfloat[Simulated urn model data. \label{fig:sim-simulated}]{%
		\includegraphics[width=0.48\columnwidth]{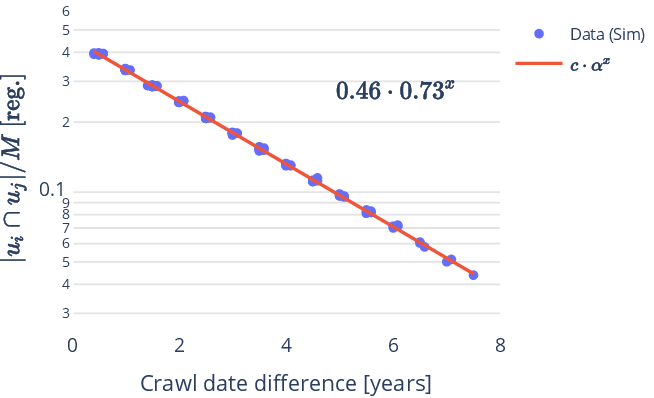}}
	\caption{Validation via simulation. (a) Regression on \GAW data. (b) Simulation with urn model parameters recovers the same relationship, confirming the method.}
	\label{fig:simulation}
\end{figure}

For the \GAW, the observed decay is in excellent agreement to our model, as depicted in Fig.~\ref{fig:simulation}(a), where we have fitted Eq.~\ref{eq:model1} via ordinary least squares regression to the log-transformed containment values ($R^2=0.95$).
In particular, we find an average decay rate of $\alpha \approx 0.73$ per year, which corresponds to approximately 27\% annual URL turnover, consistent with prior studies of web persistence~\cite{koehler2002web}. 
Most importantly, we estimate coverage as  $0.46$, i.e., an average of 46\% of the complete crawlable academic web are captured. 
{To demonstrate the validity of the analytical expression of the urn model, we compare Eq.~\ref{eq:model1} with stochastic simulations of the model in Fig.~\ref{fig:simulation}, which show perfect agreement.}

\begin{figure}[t]
	\centering
	\subfloat[Individual regression lines per starting crawl.\label{fig:percrawl-lines}]{%
		\includegraphics[width=0.48\columnwidth]{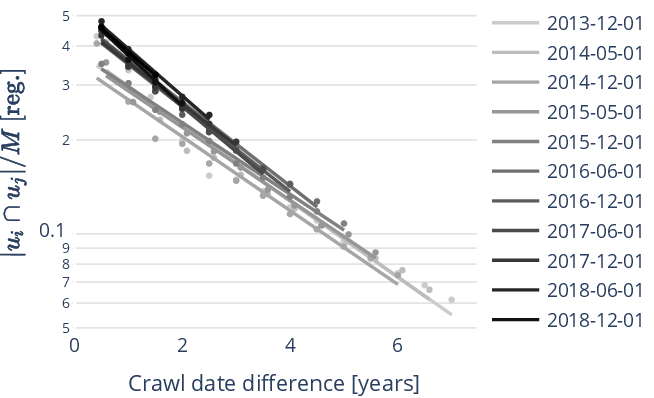}}
	\hfill
	\subfloat[Estimated $c$ vs.\ starting crawl date.\label{fig:percrawl-intercept}]{%
		\includegraphics[width=0.48\columnwidth]{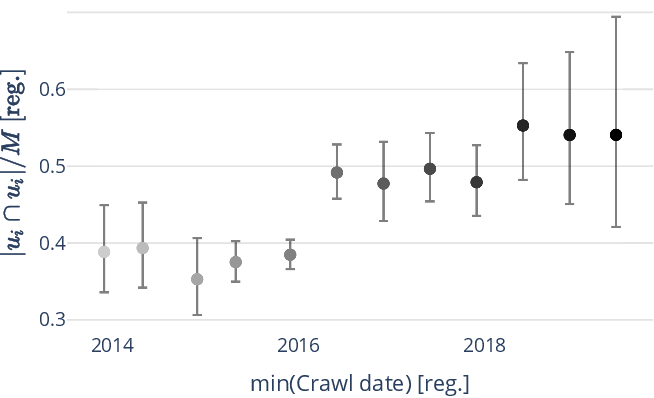}}
	\caption{Per-crawl regression analysis. (a) Separate regressions for each starting crawl yield approximately parallel lines. (b) The y-intercept (estimated $c$) shows a rising trend over 2013--2021.}
	\label{fig:percrawl}
\end{figure}

Figure~\ref{fig:percrawl} examines whether coverage varies over time by fitting separate regressions for each crawl's comparisons with all subsequent crawls.
Panel \ref{fig:percrawl-lines} depicts the individual regression lines and show consistent decay behavior with approximately parallel slopes.
Panel \ref{fig:percrawl-intercept}  presents the y-intercepts estimating the sampling fraction $c$ for each starting crawl.
This shows a rising trend over time in, indicating that crawl coverage has improved over the observation period. 
This improvement in coverage is either driven by an increase in crawl efficiency or a reduction in the crawlable population size $N$.

\section{Discussion}\label{sec:discussion}
Estimating web crawl coverage can be very challenging and constrained by the way web data is crawled. 
We presented the first method to estimate absolute web crawl coverage from longitudinal data alone, i.e., from a series of subsequent crawls.
By analyzing pairwise URL set intersections over time across crawls we were able to estimate the coverage of the crawls.
A simple urn model reproduces key features of the empirical data and therefore provides theoretical grounding for our ability to estimate coverage from the longitudinal GAW crawl data.

The key insight is that temporal variation in a longitudinal archive of web crawls serves as a substitute for the multiple independent samples required by traditional capture-recapture methods.
In particular, by observing how URL sets overlap across time, the urn model allows us to decompose the (i) coverage (y-intercept of the fitted log relationship $c$ (i.e. what fraction of the total we capture), and the (ii) decay component $\alpha$, slope of the log relationship, i.e.,  how quickly content turns over.
As a case study, we use our method to estimate our coverage of the German Academic Web. 
We find that each crawl captures approximately 46\% of the complete crawlable academic web, with content persistence of approximately 73\% per year (27\% annual URL turnover).

Our work extends \citet{bharat1998technique} from cross-sectional comparison of search engine indices to longitudinal analysis of crawled URL sets.
Where they probed multiple search engines through query interfaces and could only compute relative sizes, we extract absolute coverage from direct temporal self-intersections within a single crawl archive.
Furthermore, our extracted decay parameter $\alpha \approx 0.73$ per year—representing the annual URL persistence rate-implies a half-life of approximately 2.2 years for URL survival, closely aligning with \citet{koehler2002web}'s longitudinal finding of approximately 2-year half-lives for web page persistence.

The method enables ongoing assessment of crawl coverage without external resources. 
Specifically, it provides archive operators a self-contained method to assess coverage without knowledge of external ground truth datasets, content extraction or entity matching, and most importantly, without the ability to compare to other crawlers. 
Researchers may benefit by being able to estimate what fraction of the target web it captures, informing confidence in downstream tasks i.e.  training LLMs. 
In addition, the temporal stability analysis ({Fig.~\ref{fig:percrawl-intercept}} in Section~\ref{sec:results}) reveals whether the target population is in a ``replacement regime'' (stable size) or ``growth regime'' (expanding), guiding crawl frequency decisions.

The way a crawl is modeled here is independent of the type of flux in the URL space, the flux of the web only affects how we may fit the overlap data.
Corresponding the y-intercept to the crawl size normalized self-intersection is independent of the flux.
If the \cGAW grows over time, the sampling fraction $c_t = M/N_t$ decreases.
We would observe the coverage estimate declining.
Conversely, if $N_t$ shrinks, coverage estimates would increase.
The coverage estimate across starting years provides an internal test for dynamics of $N_t$. 
In the case of the GAW, with a non-revisit policy, the web is crawled outwards, exploring further regions without resampling previously seen URLs.
The sampling bias is accounted for through the log residual, providing a measure for an intersections deviation from the urn model.

Like any inference based on indirect observation, our approach rests on a set of assumptions and data requirements that delimit its applicability. 
We therefore discuss key limitations and boundary conditions below. 
The method assumes crawls sample approximately uniformly from the \cGAW.
Real crawls exhibit biases: breadth-first traversal favors well-connected pages; scope restrictions exclude content outside seed domains.
The coverage estimate should be interpreted as coverage weighted by the crawl configuration, not uniform coverage of all theoretically reachable URLs. 
The approach further requires multiple crawls over time.
It cannot assess coverage of a single snapshot.
Archives with a small number of crawls may have insufficient data for reliable inference of the crawl coverage.
We also assume the \cGAW size remains approximately constant between start and termination of a crawl.
In rapidly growing domains (e.g., social media), this assumption may not hold, and the intercept interpretation would be confounded by population change.
Finally, coverage is measured at the URL level, not content level.
A URL may persist while its content changes substantially.
Content-based similarity measures could complement URL intersection analysis.

Several extensions merit investigation. 
First, the urn model could be generalized to account for non-uniform sampling by incorporating weights that reflect crawl bias toward highly connected or frequently updated pages (i.e., a network-based approach). 
Second, the method could be applied to other longitudinal web archives, such as Common Crawl, or domain-specific collections, to assess its generality across crawl regimes. 
Third, URL-level intersection could be complemented with content-based similarity measures (e.g., MinHash) to estimate semantic rather than purely URL-level coverage. Finally, future work could develop diagnostics to disentangle true population growth from declining coverage in longitudinal archives.

In conclusion, we introduced a self-contained method to estimate absolute web crawl coverage using longitudinal data alone, leveraging temporal self-intersections of URL sets as a substitute for independent samples. 
By modeling overlap dynamics with a simple urn process, we showed that the intercept of the empirically observed decay in containment directly recovers the sampling fraction, while the slope captures content turnover. 
Applied to the German Academic Web, the method indicates that individual crawls capture approximately 46\% of the crawlable URL space, with an annual persistence rate of about 73\%. 
Because the approach requires neither external ground truth nor comparison across crawlers, it provides archive operators and researchers with an interpretable and general tool for assessing crawl completeness in longitudinal web archives.




\paragraph{Data availability.}
Metadata for the German Academic Web (URLs and timestamps) is available at \url{https://german-academic-web.de/}.

\section*{Acknowledgments}
The author thanks Robert J\"aschke for conceiving and maintaining the German Academic Web archive, providing access to the crawl data, and offering valuable feedback on this work.

Parts of this research were funded by the German Federal Ministry of Education and Research (BMBF) in the REGIO project (grant no.\ 01PU17012D).

\bibliographystyle{plainnat}
\bibliography{bibliography}

\end{document}